\documentclass[12pt]{amsart}

\usepackage{bbm}
\usepackage{hyperref}

\newcommand {\ssecfont} {\normalfont}
\newcommand {\sssecfont} {\normalfont}

\newcommand {\bcdot}   {\mathbin{\hbox{\raise.4ex\hbox{\bf.}}}} % bold \cdot

\newcommand {\ssbegin}[1]
 {\refstepcounter{subsection}
 \def \secno {\gdef \secno {}{\ssecfont
\thesubsection.\hskip 2ex}%
 }%
 \begin{#1}}

\newcommand {\sssbegin}[1]
 {\refstepcounter{subsubsection}
 \def \secno {\gdef \secno {}{\sssecfont
\thesubsubsection.\hskip 2ex}%
 }%
 \begin{#1}}

\newcommand{\ev}{\bar 0}
\newcommand{\od}{\bar 1}
\newcommand{\One}{{\mathbbm{1}}}
\newcommand{\Span}{\text{Span}}
\newcommand{\tr}{\text{tr}}

\def\nopoint#1#2{}

\begin{document}

\title{Supertraces on queerified algebras} %  of observables in the Calogero--Moser model,\\ Vasiliev higher spin algebras,\\ algebras of complex size matrices}

\author{Dimitry Leites${}^{a}$, Irina~Shchepochkina${}^b$}

\address{${}^a$Department of mathematics, Stockholm University, Roslagsv. 101, 
Stockholm, Sweden\\ dimleites@gmail.com\\
${}^b$Independent University of Moscow,
B. Vlasievsky per., d. 11, RU-119 002 Moscow, Russia\\
irina@mccme.ru}

%\begin{history}
%\received{(??, 2009)} \revised{(??, 2009)} \accepted{(??, 2009)}
%\comby{(xxxxxxxxx)}
%\end{history}

%\date{\today}

%--------------------------------------------------------------------
%--------------------------------------------------------------------
%--------------------------------------------------------------------
\begin{abstract} We describe supertraces on ``queerifications'' (see arxiv:2203.06917) of the algebras of matrices of ``complex size'', algebras of observables of Calogero-Moser model, Vasiliev higher spin algebras, and (super)algebras of pseudo-differential operators. In the latter case, the supertraces establish complete integrability of the analogs of Euler equations to be written.
\end{abstract}



\maketitle

%\iffalse
\markboth{\itshape Dimitry Leites\textup{,} Irina Shchepochkina} {{\itshape Traces and supertraces}}
%\fi

\thispagestyle{empty}

The traces on Lie algebras, and even (not odd) supertraces on Lie superalgebras, are  known to be very useful, for example, in representation theory, see, e.g., \cite{DSB} and references therein. The \textit{odd} supertraces are  less known, hence less popular. In Section~\ref{intinv}, we show one of their usages not previously explored: application to the study of integrability of certain hamiltonian dynamical systems.

Inspired by \cite{BLLS}, where Lie algebras are queerified over an algebraically closed ground field of characteristic $p=2$ in order to produce the complete list of simple finite-dimensional Lie superalgebras in characteristic $p=2$, this new method --- queerification --- producing many new simple Lie superalgebras from associative algebras and superalgebras is applied in \cite{L}  over $\mathbb{C}$ to several infinite-dimensional algebras of interest in theoretical physics. A~ number of papers were devoted to description of traces on these algebras, and supertraces on these algebras considered as superalgebras, see \cite{KS, KT1, KT2, KT3}. In this note, we describe the supertraces on the Lie queerifications of these algebras and superalgebras, having added one more type of examples.

\section{Preliminaries}\label{S1}

\subsection{From associative to Lie} Let $\mathbb{K}$ be an algebraically closed ground field of characteristic $p\neq 2$; unless otherwise stated, we consider $\mathbb{K}=\mathbb{C}$. 

Let $A$ be any associative algebra, let $A^L$ be the Lie algebra whose space is $A$ but multiplication is given by the commutator $[a,b]:=ab-ba$ for any $a,b\in A$. 

Let $A$ be a~$\mathbb{Z}/2$-graded algebra; let $p$ denote the parity function. If $A$ a~$\mathbb{Z}/2$-graded  associative algebra, let $A^S$ be the Lie superalgebra whose space is $A$ but the multiplication is given by the supercommutator, which by the modern habitual abuse of notation is also denoted $[a,b]$ although defined  to be $[a,b]:=ab-(-1)^{p(a)p(b)}ba$. Let $\mathfrak{g}':=[\mathfrak{g},\mathfrak{g}]$ be the first derived Lie algebra (resp. Lie superalgebra), a.k.a. commutant  (resp. supercommutant), of the Lie algebra (resp. Lie superalgebra) $\mathfrak{g}$. 

Not every $\mathbb{Z}/2$-graded algebra $\mathcal{A}$ is called \textit{superalgebra}: only if  multiplication in $\mathcal{A}$ or (if $\mathcal{A}$ is associative) in $\mathcal{A}^S$ depends on the parity. 

Recall, that  a \textit{trace} (called \textit{supertrace}  in the super setting, for emphasis) on a given Liealgebra (resp. Lie superalgebra) $\mathfrak{g}$ is  a linear function that vanishes on its commutant (resp. supercommutant), so there are $\dim (\mathfrak{g}/[\mathfrak{g},\mathfrak{g}])$  linearly independent traces (resp. supertraces) on~ $\mathfrak{g}$, some of the supertraces can even and some of them odd.

\subsection{Queerifications in characteristic  $p\neq 2$ (from \cite{BLLS})}\label{111} Let $A$ be an
associative algebra.  The space of the associative algebra $\text{Q}(A)$ --- the \textit{\textbf{associative} queerification} of $A$ --- is $A \oplus\Pi(A)$, where
$\Pi$ is the change of parity functor, with the same multiplication in $A$ and the adjoint action of  $A$ on $\Pi(A)$; let 
\[
\Pi(x)\Pi(y):= xy\text{~for any~~}x,y\in A.
\]
Set $\text{Q}(n):=\text{Q}(\text{Mat}(n))$. 

We will be mostly interested in the following \textit{\textit{\textbf{Lie}} queerifications} $\mathfrak{q}(A)$:  

1) for $A$ an associative algebra;

2) for $A$ a $\mathbb{Z}/2$-graded associative algebra (\lq\lq Liefication" yields $\mathbb{Z}/2$-graded Lie algebra $A^L$);

3) for $A$ an associative superalgebra (this case differs from case 2) because passing to  $A^S$ the supercommutators instead of commutators are considered; \lq\lq super Liefication" yields Lie superalgebra $A^S$ which is $\mathbb{Z}/2$-graded  by parity. %; considered as algebra it is $(\mathbb{Z}/2\times\mathbb{Z}/2)$-graded).

In case 1), as spaces, $\mathfrak{q}(A):={A^L\oplus\Pi(A)}$, so $\mathfrak{q}(A)_{\bar 0}=A^L$ and $\mathfrak{q}(A)_{\bar 1}=\Pi(A)$, with the bracket given
by the following expressions and super anti-symmetry, i.e., anti-symmetry amended by the Sign Rule:
\begin{equation}\label{commRel}
{}[x,y]:=xy-yx;\quad [x,\Pi(y)]:=\Pi(xy-yx);\quad [\Pi(x),\Pi(y)]:=
xy+yx \ \text{~for any~~}x,y\in A.
\end{equation}

The term ``queer", now conventional, is taken after the Lie
superalgebra $\mathfrak{q}(n):=\mathfrak{q}(\text{Mat}(n))$. (The associative superalgebra $\text{Q}(n)$ is an analog of $\text{Mat}(n)$; likewise, the Lie superalgebra $\mathfrak{q}(n)$ is an analog of $\mathfrak{gl}(n)$  for several reasons, for example, due the role of these analogs in Schur's Lemma and in the classification of central simple superalgebras, see \cite[Ch.7]{Lsos}.) We express the elements
of the Lie superalgebra $\mathfrak{q}(n)$ by means of a~pair of matrices
\begin{equation}
\label{.1} (X,Y)\longleftrightarrow
\begin{pmatrix}X&Y\\Y&X
\end{pmatrix}
\in \mathfrak{gl}(n|n), \text{~~where $X,Y \in \text{Mat}(n)$}.
\end{equation}

For any associative $A$, we will similarly denote the elements of $\mathfrak{q}(A)$ by pairs $(X,Y)$, where $X,Y \in A$. The brackets between these elements are as follows:
\begin{equation}\label{.2}
\renewcommand{\arraystretch}{1.4}
\begin{array}{l}
{}[(X_1,0),(X_2,0)]:=([X_1,X_2],0),\quad
[(X,0),(0,Y)]:=(0,[X,Y]),\\
{}[(0,Y_1), (0,Y_2)]:=(Y_1Y_2+Y_2Y_1,0).
\end{array}\end{equation}

We define Lie queerifications in cases 2) and 3) in the next section.

\section{Traces and supertraces: generalities}{}~{}

\ssbegin{Theorem} Let $A$ be an associative algebra. Then, the estimate of the number of traces in the $3$ cases  of its Lie queerification are described below. \end{Theorem}

\subsubsection{Case $1$}\label{2.1}
Let $A$ be an associative algebra, then $Q(A):=A\oplus \Pi(A)$ is the \textit{associative queerification} of $A$; let $\mathfrak{g}:=A^L$ be the corresponding Lie algebra and let the Lie superalgebra $\mathfrak{q}\mathfrak{g}:=(Q(A))^S=\mathfrak{q}(A)$ be the \textit{Lie queerification} of $A$.
Clearly, $\mathfrak{q}\mathfrak{g}=\mathfrak{g}\oplus \Pi(\mathfrak{g})$ as spaces.

Denote, for brevity, $\mathfrak{u}:=(\mathfrak{q}\mathfrak{g})'$. By definition the supercommutant $\mathfrak{u}:=\mathfrak{u}_{\ev}\oplus \mathfrak{u}_{\od}$ is spanned by elements $[a,b]$ for any $a,b\in \mathfrak{q}\mathfrak{g}$. In particular, $\mathfrak{u}_{\od}$ is spanned by elements of the form $[x,\Pi(y)]=\Pi([x,y])$ for any $x,y\in  (\mathfrak{q}\mathfrak{g})_{\ev}$, and hence $\mathfrak{u}_{\od}=\Pi([\mathfrak{g},\mathfrak{g}])$. Therefore, there are as many odd supertraces on $\mathfrak{q}\mathfrak{g}$ as there are traces on $\mathfrak{g}$.

Clearly, $\mathfrak{u}_{\ev}$ is the sum of two ideals of $\mathfrak{g}$:
\[
\mathfrak{u}_{\ev}=[\mathfrak{g},\mathfrak{g}]+[\Pi(\mathfrak{g}),\Pi(\mathfrak{g})].
\]
Observe that the second summand does not have to be contained in the first one. Therefore, there are fewer even supertraces on $\mathfrak{q}\mathfrak{g}$ than there are traces on $\mathfrak{g}$.

In particular, if  $A$ has unit $\One$, 
%and $\mathfrak{g}=\mathbb{K}\cdot \One\oplus \mathfrak{g}'$,  
then $\mathfrak{u}_{\ev}=\mathfrak{g}$, because $[\Pi(\One),\Pi(x)]=2x$ for any $x\in \mathfrak{g}_{\ev}$.

For example, if $A=\text{Mat}(n)$, then on $Q(A)$ there is an odd trace, nowadays called \textit{queertrace}; it was first defined in \cite{BL} by the formula
\[
\text{qtr}: \begin{pmatrix}X&Y\\Y&X
\end{pmatrix}\mapsto \tr Y.
\]

\subsubsection{Case $2$}\label{2.3}
Let $A$ be a $\mathbb{Z}/2$-graded associative algebra, $\mathfrak{g}:=A^L$. Clearly, $\mathbb{Z}/2$-grading of~ $A$ makes $\mathfrak{g}=\mathfrak{g}_{\ev}\oplus \mathfrak{g}_{\od}$ a~ $\mathbb{Z}/2$-graded Lie algebra.

Now, consider $A$ as an associative superalegbra; let $\mathfrak{s}=\mathfrak{s}_{\ev}\oplus \mathfrak{s}_{\od}:=A^S$. Clearly, $\mathfrak{s}_{\ev}=\mathfrak{g}_{\ev}$ and $\mathfrak{s}_{\od}=\Pi(\mathfrak{g}_{\od})$, as spaces. The first equality is, moreover, true as an isomorphism of Lie algebras, whereas the second one as an isomorphism of  $\mathfrak{s}_{\ev}=\mathfrak{g}_{\ev}$-modules. 

The supercommutants are as follows:
\[
\begin{tabular}{|l|}
\hline
$\mathfrak{g}'=(\mathfrak{g}')_{\ev}\oplus (\mathfrak{g}')_{\od}$,  where for any $x,y\in\mathfrak{g}_{\ev}$ and $a,b\in\mathfrak{g}_{\od}$, we have\\ 
$(\mathfrak{g}')_{\ev}=\Span([x,y]=xy-yx,[a,b]=ab-ba),\ (\mathfrak{g}')_{\od}=\Span([x,a]=xa-ax)$;\\
\hline
$\mathfrak{s}'=(\mathfrak{s}')_{\ev}\oplus (\mathfrak{s}')_{\od}$,  where for any  $x,y\in\mathfrak{s}_{\ev}$, $a,b\in\mathfrak{s}_{\od}$ we have\\
$(\mathfrak{s}')_{\ev}=\Span([x,y]=xy-yx,[a,b]=ab+ba), \ (\mathfrak{s}')_{\od}=\Span([x,a]=xa-ax)$.\\
\hline
\end{tabular}
\]

Therefore, $\mathfrak{s}'_{\od}=\Pi((\mathfrak{g}')_{\od})$, but \text{a priori}  we can not say anything about relations between $(\mathfrak{g}')_{\ev}$ and $(\mathfrak{s}')_{\ev}$, except for both contain $(\mathfrak{g}_{\ev})'=(\mathfrak{s}_{\ev})'$.

\subsubsection{Case $3$}\label{2.2}
Now, let $A=A_{\ev}\oplus A_{\od}$ be an associative superalgebra, $\mathfrak{g}=\mathfrak{g}_{\ev}\oplus \mathfrak{g}_{\od}:=A^S$. 

Let $Q(A):=A\oplus \Pi(A)$ and $\mathfrak{q}\mathfrak{g}:=(Q(A))^S$; clearly, $\mathfrak{q}\mathfrak{g}=\mathfrak{g}\oplus \Pi(\mathfrak{g})$ as superspaces.
It is also clear that the natural $(\mathbb{Z}/2\times \mathbb{Z}/2)$-grading on $Q(A)$  induces the same grading on  $\mathfrak{q}\mathfrak{g}$, where
$$
(\mathfrak{q}\mathfrak{g})_{\ev,\ev}=\mathfrak{g}_{\ev}, \; (\mathfrak{q}\mathfrak{g})_{\ev,\od}=\mathfrak{g}_{\od}, \; (\mathfrak{q}\mathfrak{g})_{\od,\ev}=\Pi(\mathfrak{g}_{\od}), \; (\mathfrak{q}\mathfrak{g})_{\od,\od}=\Pi(\mathfrak{g}_{\ev}).
$$

Let us now compare the supercommutants of $\mathfrak{g}$ and $\mathfrak{q}\mathfrak{g}$. We will denote the elements of $\mathfrak{g}_{\ev}$ by letters $x, y,\dots$, the elements of $\mathfrak{g}_{\od}$ by letters $a,b,c,\dots$.
We have $\mathfrak{g}'=(\mathfrak{g}')_{\ev}\oplus (\mathfrak{g}')_{\od}$,  where 
$$
 (\mathfrak{g}')_{\ev}=\Span([x,y]:=xy-yx, \ \ [a,b]:=ab+ba), \ \ (\mathfrak{g}')_{\od}=\Span([x,a]:=xa-ax).
$$

The elements that span homogeneous components of $\mathfrak{q}\mathfrak{g}$ are as follows:
\[
\begin{tabular}{|c|c|c|c|}
\hline
$\mathfrak{u}_{\ev,\ev}$ & $\mathfrak{u}_{\ev,\od}$ & $\mathfrak{u}_{\od,\ev}$ & $\mathfrak{u}_{\od,\od}$\\
\hline
$[x,y]=xy-yx$ & $[x,a]$ & $[x,\Pi(a)]$ & $[x,\Pi(y)]$\\
$[a,b]=ab+ba$ &$=xa-ax$ &$=\Pi(xa-ax)$ &$=\Pi(xy-yx)$\\
$[\Pi(a),\Pi(b)]=ab-ba$ & $[\Pi(a),\Pi(x)]$ & $[a,\Pi(x)]$ & $[a,\Pi(b)]$\\
$[\Pi(x),\Pi(y)]=xy+yx$ &$=ax-xa$ &$=\Pi(ax+xa)$ &$=\Pi(ab-ba)$ \\
\hline
\end{tabular}
\]

Therefore, $\mathfrak{u}_{\ev,\ev}=(A_{\ev})^2+(A_{\od})^2$ and $\mathfrak{u}_{\od,\ev}=\Pi(A_{\ev} \cdot A_{\od})$, as spaces. In particular, if $A$ has unit,  then 
\[
\text{$\mathfrak{u}_{\ev,\ev}=(\mathfrak{q}\mathfrak{g})_{\ev,\ev}=\mathfrak{g}_{\ev}$ and $\mathfrak{u}_{\od,\ev}=(\mathfrak{q}\mathfrak{g})_{\od,\ev}=\Pi(\mathfrak{g}_{\od})$.}
\]

In general, \text{a priori} we can only say that 
$\mathfrak{u}_{\ev,\od}=(\mathfrak{g}')_{\od}$.

\section{Examples of supertraces on queerified algebras and superalgebras}

\subsection{Clifford--Weyl algebras and superalgebras}\label{3.1} Vasiliev was, most probably, the first to publish that \textit{on the Weyl  algebra $W_n$ of polynomial differential operators in $n$ even indeterminates $x_i$ considered as superalgebra with parity given by $p(x_i)=p(\partial_i)=\od$ for all $i$, there is an even supertrace}, see \cite{Va1}. For a generalization to the Calogero-Moser case, see \cite{Va2}; the detailed version \cite{Va3} contains an elementary proof of uniqueness of the (super)trace; the construction yields algebras of \lq\lq matrices of complex size" which first appeared as associative algebras in the book \cite{Di} and as Lie algebras in \cite{Fei}.
For a~ generalization to symplectic reflection algebras  in \cite[Th.7.1.1]{KT1}. Alexey Lebedev suggested a~ beautiful elementary proof of the existence of the trace on $W_n$, see \S\ref{sL}. 

Obviously unaware of Vasiliev's works, his results were rediscovered by mathematicians, see \cite[Proposition~4.3]{Mon} and \cite{Mus}. Montgomery found out the following sufficient condition to the super version of Herstein's theorem and formulated it in the infinite-dimensional situation (in the finite-dimensional case, it is also true). 

\sssbegin{Theorem}{\textup{(\cite[Th.3.8]{Mon})}}\label{ThM} Let $A$ be a~$\mathbb{Z}/2$-graded simple associative algebra of characteristic $p\neq 2$ with supercenter $Z$ whose elements supercommute with any $a\in A$. Let the condition
\begin{equation}\label{Cond}
\text{if $u^2 \in Z$, then $u\in Z$ for any homogeneous $u\in A_{\bar 1}$}
\end{equation} 
hold. Then, $SL(A):=(A^S)^{(1)}/((A^S)^{(1)}\cap Z)$ 
is a~simple Lie super algebra.\end{Theorem}

Observe that the superalgebra $A$ of differential operators in any finite number of  indigenously odd indeterminates (a.k.a. the \textit{Clifford} algebra on $2n$ generators considered as a $\mathbb{Z}/2$-graded associative superalgebra) is isomorphic to the matrix superalgebra $\text{Mat}(2^{n-1}|2^{n-1})$ on which there is an even supertrace, whereas on $\mathfrak{q}(2^{n}):=\mathfrak{q}(\mathfrak{gl}(2^{n-1}|2^{n-1}))$ there is the (well-known today) odd \textit{queer trace}.
Therefore, by arguments in Subsection~\ref{2.3}, \textbf{there are no even supertraces on $\mathfrak{q}A$, but there is one odd supertrace}.

\subsection{(Super)algebras of ``matrices of complex size''}\label{3.2} Absolutely the same arguments as in Example \ref{3.1} are applicable to the queerifications of algebras and superalgebras $A$ of ``matrices of complex size'', see \cite[Subsection 2.2]{L} with the same  answer. (Although we do not need it here, note --- just for its beauty --- that since $\mathfrak{gl}(\lambda)=\mathfrak{gl}(\lambda)'\oplus \mathbb{C}\,1$, one can define the trace on the Lie algebra $\mathfrak{gl}(\lambda)$ by any value on~ $1$. J.~Bernstein defined the trace on $\mathfrak{gl}(\lambda)$ for $\lambda\in\mathbb{C}\setminus\mathbb{Z}$, see \cite{KhM}, so that $\tr(1)=\lambda$; Bernstein's trace naturally generalizes the trace on $\text{Mat}(|n|)$  for any $n\in\mathbb{Z}\setminus\{0\}$ such that $\tr(1_{|n|})=|n|$.)

\subsection{Symplectic reflection algebras and superalgebras}\label{3.3} For the classification of traces (resp. supertraces) on these algebras and superalgebras $A$, see \cite[Tables on pp.5,6]{KS}. Considering them as algebras (resp. superalgebras) we get the exact number of supertraces on their Lie qeerifications, according to general results established in Case \ref{2.1} (resp. Case \ref{2.3}).

For a description of ideals in these algebras and superalgebras, see \cite{KT1,KT2, KT3}. It is interesting to determine when these ideals are themselves simple algebras and superalgebras, and describe (super)traces on them and on their queerifications.

\subsection{(Super)trace on the (super)algebra of pseudo-differential operators}\label{SSpseu}
 
 \subsubsection{$N=0$} Recall that the associative algebra $\Psi$ of pseudo-differential operators of integer order is $\mathcal{F}((D^{-1}))$, where $D:=\frac{d}{dx}$  and $\mathcal{F}$ is the algebra of functions in $x$, with multiplication given for any integer $n$ by the Leibniz rule 
 \[
 D^{n}f:=\sum_{k\geq 0}\binom{n}{k}D^{k}(f)D^{n-k}, \text{~~where  $\binom{n}{k}:=\frac{n(n-1)\dots(n-k+1)}{k!}$, for any $f\in\mathcal{F}$}.
 \]
Adler defined a~ trace on the algebra $\Psi$ of pseudo-differential operators, see \cite{A} and a very reader-friendly reviews \cite{R, Ma}, as the composition of the residue and the indefinite integral
\[
\tr(\sum_{k\leq n}f_kD^{k})=\int f_{-1}dx, \text{~~where $f_k\in\mathcal{F}$}. 
\]
This trace (it vanishes on the commutators even before the integral is taken, just residue suffices, see \cite{R}) takes values in $\mathcal{F}$. According to Subsection~\ref{2.3}, \textbf{there are no even supertraces on $\mathfrak{q}\Psi$, but there are $\geq 1$ odd supertraces}; we conjecture there is just one odd supertrace.
 
\subsubsection{$N=1$}\label{ss3.4.2} On the superalgebra $\Psi_{1}:=\mathcal{F}((D^{-1}))$ of $N=1$-extended pseudo-differential operators, where $D:=\frac{\partial}{\partial \xi}+\xi\frac{\partial}{\partial x}$ and $\mathcal{F}$ is the algebra of functions in the even $x$ and odd~ $\xi$, Manin and Radul defined an even supertrace, see \cite{MR}. According to Subsection~\ref{2.3}, \textbf{there are no even supertraces on $\mathfrak{q}\Psi_{1}$, but there are $\geq 1$ odd supertraces}; we conjecture there is just one odd supertrace.

\section{An application of traces: integrals in involution}\label{intinv} Let $A$ be an associative (super)algebra, $\tr$ a (super)trace on~ $A$, and  $b$ the corresponding  \textit{invariant symmetric} bilinear form (briefly: IS form)
\[
b(X, Y):=\tr(XY)\text{~~for any $X, Y\in A$}. 
\]
Let, moreover, $b$ be non-degenerate, briefly: NIS. 

Let $\tr$, and hence $b$, be even. Then, for the equation ($L$ and $P$ are in honor of Peter Lax)
\begin{equation}\label{Lax}
\dot L=[L, P], \text{~~where $L, P\in A$ and dot signifies derivative with respect to time $t$},
\end{equation}
and for its more subtle version, more adequate in the super setting with $1|1$-dimensional Time with even coordinate $t$ and odd one $\tau$, see \cite{Sha}:
\begin{equation}\label{11Lax}
(\partial_\tau+\tau\partial_t) L=[L, H], \text{~~where $L, H\in A$},
\end{equation}
the functions $L\mapsto \tr(L^k)$ are integrals in involution with respect to the Poisson bracket defined on the space of functions on $\mathfrak{g}^*$ as follows, see, e.g., \cite{A}. We identify $\mathfrak{g}$ with the linear functions on $\mathfrak{g}^*$ and for any functions $f,g\in\mathcal{F}(\mathfrak{g}^*)$ the Poisson bracket is defined to be
\begin{equation}\label{PoiB}
\{f,g\}(X):=X([df(X),dg(X)])\text{~~for any $X\in\mathfrak{g}^*$}.
\end{equation}

It seems, nobody considered yet the Euler equations or Lax pairs \eqref{Lax} related with superalgebras $A$ on which there is an odd trace $\text{qtr}$, and hence an odd $b$. If $b$ is odd, then $\mathfrak{g}\simeq\Pi\mathfrak{g}^*$, and an antibracket, rather than a~Poisson bracket, is defined on the space $\Pi\mathcal{F}(\mathfrak{g}^*)$ of functions on $\mathfrak{g}^*$. The functions $L\mapsto\text{qtr}(L^k)$ are integrals in involution, i.e., commuting with the Hamiltonian and each other (themselves including in the case of the anti-bracket) with respect to the antibracket.

On $2n|k$-dimensional superspace on which the Poisson bracket is defined or $n|n$-dimensional superspace  which the anti-bracket bracket is defined, let there be $n$ first integrals of the dynamical system~ \eqref{Lax} in involution. Then, a theorem of Shander guarantees complete integrability of the system whatever $k<\infty$ is, see \cite{Sh}. 

The traces on (super)algebras $A$ considered in Subsection~\ref{SSpseu} determine what researchers conceded to call ``complete integrability" in the case of infinite-dimensional Hamiltonian system (since there are infinitely many of these traces --- \lq\lq this infinity is a half of the infinite dimension"). Examples: the KdV equations (resp. $(N=1)$-extended KdV, see Subsection~\ref{ss3.4.2}) for the case where $L$ is the Schr\"odinger operator (resp. its $(N=1)$-extension). 

\textbf{Open problem}: For any simple associative algebra $A$, give examples of integrable systems related with $Q(A)$.

\section{Supertrace on $W_n$}\label{sL}

\subsection{Two general facts (\cite{S})}\label{GF}

1) If there is no (super)trace on an associative (super)algebra $A$, then there is no (super)trace on any product $A\otimes B$ for any associative (super)algebra $B$ with unit: if $[a_1, a_2] = a$, then  $[a_1\otimes 1, a_2\otimes b] = a\otimes b$ for any $a_1, a_2\in A$ and $ b\in B$, i.e., if any element of  $A$ can be represented as a linear combination of (super)commutators, then the same is true for any element of  $A\otimes B$.
 
2) Let $\tr_i$ be a  (super)trace on the associative (super)algebra $A_i$ for $i=1,2$. Then,  
\[
\text{$\tr (a_1\otimes a_2) := (\tr_1\, a_1)(\tr_2\,  a_2)$ for any $a_i\in A_i$}
\]
 is a (super)trace on $A_1\otimes A_2$.

\ssbegin{Lemma}\label{Le} Consider $W_n$ as a superalgebra with $p(x_i)=p(\partial_{x_i})=\od$. 

Then, on $W_n$, there is an even trace.\end{Lemma}

Observe that if we consider $W_n$ as an algebra, not a~superalgebra, no analog of Lemma~\ref{Le} takes place since the associative algebra $W_n$ is simple; on the other hand, the center (constants) is given by a non-trivial cocycle.

\begin{proof}(A. Lebedev) Actually, $W_n=\mathbb{K}1\oplus [W_n, W_n]$, where $[W_n, W_n]$ is the supercommutant. The proof below demonstrates existence of the trace, but its uniqueness (up to a~non-zero factor) should be proved separately. For the proof of uniqueness, see \cite{Va3, Mon}.

Let $n=1$. Let $d:=\frac{d}{dx}$, let the \text{weight} $\text{wht}$ of $x$ be equal to $1$, let $\text{wht}(d):=-1$. On $W_n$, define the following linear function $T$:
\[
T(P) :=\begin{cases} (P(\frac{1}{x+1}))|_{x=1}&\text{ if $\text{wht}(P)=0$,}\\
0&\text{ if $\text{wht}(P)\neq 0$}.
\end{cases}
\]
Let us prove that $T$ is a supertrace on $W_n$, i.e.,
\[
T(PQ) = (-1)^{p(P)p(Q)}T(QP). 
\]
Clearly, it suffices to prove this for the case where $P$ and $Q$ are monomials whose weights are opposite.
 
Case 1: $P= x^{n+1}d^n$ and $Q=d$. Then,
\[
\begin{array}{ll}
T(PQ) &= (x^{n+1} (-1)^{n+1} \frac{(n+1)!}{(x+1)^{n+2}})|_{x=1} = -(-\frac12)^{n+2} (n+1)! \ ,\\ 
T(QP) &= (d (x^{n+1} (-1)^n \frac{n! }{(x+1)^{n+1}})|_{x=1} \\
&= ( (-1)^n \frac{(n+1)! \ x^n }{(x+1)^{n+1}} - (-1)^n \frac{(n+1)! \ x^{n+1}}{(x+1)^{n+2}})|_{x=1} = (-\frac12)^{n+2} (n+1)!
\end{array}
\]
 
This implies the answer for the case where $Q=d$ and any $P$ of weight $1$, because any such $P$ can be represented as a linear combination of operators of the form $x^{n+1}d^n$.
 
Case 2: $\text{wht}(P)=-1$ and $Q=x$. Then,
\[
\begin{array}{l}
T(QP) = (xP(\frac{1}{x+1}))|_{x=1} = (P(\frac{1}{x+1}))|_{x=1},\\
T(PQ) = (P(\frac{x}{x+1}))|_{x=1} = (P(1-\frac{1}{x+1}))|_{x=1} = -(P(\frac{1}{x+1}))|_{x=1}, 
\end{array}\]
since $P(1)=0$ because $\text{wht}(P)<0$.
 
This implies the general case where  $P$ and $Q$ are monomials of opposite weights, because we can transplant $x$ and $d$, one by one, from the end of $PQ$ to the beginning until we get $QP$, and each transplantation changes the sign by the opposite;  by $(-1)^{deg(Q)} = (-1)^{p(P)p(Q)}$ altogether.
 
Since $T(1)=\frac12$, it follows that 1 can not be represented as a linear combination of supercommutators. 

For $n>1$, recall that $W_n\simeq W_{n-1}\otimes W_1$ and apply general fact 2), see Subsection~\ref{GF}. \end{proof} 

\textbf{Comment: how to guess the form of $T$}. Let $P$ be a differential operator of weight~ 0.  In the basis $1$, $x$, $x^2$, \dots , consider $P$ as a linear operator on the space of polynomials and consider the matrix of $P$. Naively, ignoring possible divergence, the supertrace of this matrix is equal to
\[
\sum_{n=0}^\infty (-1)^n \text{(the coefficient of $x^n$ in $Px^n$)}.
\]
Since $\text{wht}(P)=0$, then $Px^n$ is equal to the above-mentioned coefficient of $x^n$. In other words, the coefficient is equal to the value of $Px^n$ at $x=1$. Hence, the supertrace is equal to
\[
\sum_n (-1)^n Px^n|_{x=1} = P(1-x+x^2-x^3+...)|_{x=1} = \left(P\left(\frac{1}{1+x}\right)\right)|_{x=1}.
\]

\subsection*{Acknowledgements} We are thankful to A.~Lebedev for help. DL was supported by the grant AD 065 NYUAD.

%--------------------------------------------------------------------
%--------------------------------------------------------------------
%--------------------------------------------------------------------
\end{document}